\documentclass[%
 preprint,
nofootinbib,
 amsmath,amssymb,
 aps,
 pre,
floatfix,
]{revtex4-2}

\usepackage{float}
\usepackage{listings}
\usepackage{graphicx}
\usepackage{adjustbox}

\usepackage[font=small]{caption}
\usepackage[labelformat=empty, position=top]{subcaption}

\usepackage{amsmath}
\usepackage{mathtools}  
\usepackage{diffcoeff}  
\usepackage{amsthm}
\usepackage{afterpage}
\usepackage{xcolor}

\usepackage{graphicx}% Include figure files
\usepackage{dcolumn}% Align table columns on decimal point
\usepackage{bm}% bold math

\begin{document}

\title{Classification of Cellular Automata based on Statistical Mechanics}

\author{Luca Bertolani}
\author{Andrea Idini}%
 \email{andrea.idini@matfys.lth.se}
\affiliation{%
Division of Mathematical Physics, Department of Physics, LTH, \\ Lund University, PO Box 118, S-22100 Lund, Sweden}

\date{\today}

\begin{abstract}
Cellular automata are a set of computational models in discrete space that have a discrete time evolution defined by neighbourhood rules. They are used to simulate many complex systems in physics and science in general. 
%Over the years many studies have tried to characterize their behaviour within a scheme in order to classify and identify their properties. 
In this work, statistical mechanics and thermodynamics are used to analyse a large set of outer totalistic two-dimensional cellular automata. Thermodynamic variables and potentials are derived and computed according to three different approaches to determine if a cellular automaton rule is representing a system akin to the ideal gas, in or out of the thermodynamical equilibrium. It is suggested that this classification is sufficiently robust and predictive of interesting properties for particular set of rules.
\end{abstract}

\maketitle
\newpage

\section{Introduction}
\label{sec:introduction}

%\subsection{Cellular Automata}
A cellular automaton (CA) is a computational system that is composed of a set of cells defined on a lattice. Each cell has a state that evolves following discrete time--steps. The state is updated to the next time--step according to some rules which, depending on the state of the cell's neighbourhood, give the configuration at the next time step. Given this simple definition, cellular automata have demonstrated a surprising complexity and range of applications. Cellular automata are a typical example of the emergence of complex patterns and behaviours from a simple set of rules. Most literature focuses on binary cellular automata of one and two dimensions. Binary means that each cell can take one of two values, sometimes it also referred to as Boolean. In this work, we study and classify totalistic rules of two--dimensional binary cellular automata. Totalistic, or additive, means that the update of the state of the cell is described by the total number of neighbours in a given state, instead of the value of each neighbourhood site.

In general, the number of available states does not necessarily need to be two. Furthermore, the neighbourhood can be defined in different ways in arbitrary dimensions \cite{1983}.  
However, dimensions higher than two require higher computational power and have a much larger number of possible rules. Additionally, there is a greater proportion of rules that generate trivial outcomes. Similarly, the study of non-totalistic and eventually extended neighbourhood cellular automata would massively increase the number of rules, beyond the scope of this paper.

In the last few decades, there have been several studies to evaluate and characterize the complexity of cellular automata, in both qualitative and quantitative ways.
Among the qualitative evaluation of complexity emerging from different rules, it is worth mentioning Stephen Wolfram's four-class classification~\cite{1983} and the six class classification by  Li, Packard and Langton~\cite{phase}.
Quantitatively, complexity has been tackled starting from many different assumptions. Most approaches make use of entropy as a measure of complexity~\cite{1983,phase,complexity,causal}. Other works use concepts such as mutual information, causal entropic forces, self-organized criticality and 1/f noise, compression length as well as the definition of completely new quantities~\cite{complexity,evolving,boundary,causal,simon}. 

CA have been applied in different fields and contexts, specifically in systems that iterate over local rules (cf.~\cite{app,astro} and refs. therein). These include disease spreading \cite{Diseases} and recent applications to protein folding \cite{varela2022evolving} and transport phenomena \cite{Dmitri:22}. Therefore, the finding of simple reliable metrics for evaluating the behaviour of cellular automata can be very beneficial in accelerating the development of further applications. %Here metrics is intended as a standard of measurement or evaluation.
Furthermore, CA have also been studied in relation to statistical mechanics, mainly for the investigation of phase transitions~\cite{phase} and the Ising model~\cite{ising}. Since CA offer a lattice grid with two values, they can be made to correspond to spin up and spin down. Historically, cellular automata have been used to model physical systems and vice-versa, with present interest in the thermodynamic characterization of cellular automata at quasi-equilibrium \cite{Pazzona-PRE}. 

%A distinction is usually made between reversible and irreversible rules and between additive and non-additive rules (sometimes also referred to as totalistic/non-totalistic). A rule is reversible if there is a one-to-one mapping between the initial and the final state of the automaton, irreversible if different initial states produce the same final state\cite{1983}. A rule is said to be additive if the update is the result of the sum of the neighbourhood instead of the specific values at each neighbourhood site \cite{1983}.

%The originality and novelty of this approach lies in the premise of using the most basic approaches possible to study the problem of the thermodynamics of cellular automata and in deriving a quantitative classification scheme, as opposed to the most used ones to date which are qualitative.
This work studies cellular automata using different metrics from statistical mechanics and classical thermodynamics, in and out of equilibrium for all the totalistic rules that do not distinguish between the values of the central states. These metric are tested to find a way to quantitatively describe the behaviour arising from different rules. There has been work in this direction\cite{eq,stat}, with particularly relevant applications to lattice gases \cite{rothman2004lattice}, but no comprehensive study of how to translate thermodynamical macrostates to the framework of cellular automata enabling a classification involving a large scale account of totalistic rules. Most cellular automata rules are not invertible, making the application of thermodynamics unphysical, hence the correspondence with a physical system an analogy. However, despite not being representative of a straightforward interpretation of physical property, the statistical mechanics approach used in this work can provide insight with %distinguishes itself from other classification schemes, which instead derives theorems and statements about cellular automata by inspecting the system at each step in more of a 'bottom-up' fashion, which would include the investigation of the configuration matrix at each step. The former might lack the detail of the latter but 
the great advantage of processing a large number of CA rules in an automated way. The classification obtained can serve as a pre--selection of interesting rules. 

In the following section, we assume a definition for the energy of different states, and define different ways to compute the entropy, temperature, heat capacity, Helmholtz free energy, and pressure. We compare these quantities derived from a benchmark with the ideal gas definitions, in order to identify whether a particular rule is akin to an ideal gas. We also compare the benchmark to quantities based on partition function or non--equilibrium treatments. We use this to classify a particular rule accordingly. In the result section we discuss the stability of the results obtained.

%The Manuscript is organised as follows. At first, in the method section, the used notation and definitions of cellular automata will be introduced [this sentence sounds weird and manuscript is technically incorrect (because it is not a manuscript)]
% The Work is organised as follows. At first, the notation and definitions of cellular automata are introduced in the method section. 
% Then, we provide the general definitions of the thermodynamic quantities under consideration. We proceed in the classes section, with the translation of thermodynamical variables to the cellular automata framework and the effective redefinition of these variables in the different context that enables us to compare a given rule with the ideal gas or to systems in and out of the thermodynamical equilibrium.  In the result section, the computational results will be presented and finally discussed.
%Subscripts are used to distinguish the quantities derived with the different approaches, to avoid confusing the reader.
%The benchmark quantities have no subscript ($Q$), the ideal gas quantities are written as $Q_I$ and the partition function quantities are written as $Q_Z$.

\section{Classes}
 \subsection{Benchmark}
 \label{sec:bm}
Cellular automata can lend themselves to the framework of statistical mechanics \cite{1983}. The system is partitioned in cells by construction and the state of each cell $i$ can be interpreted as an unit--less energy state $E_i$. In the case of binary cellular automata, this leads to two possible energy states per cell, namely $E_i=1$ and $E_i=0$. We consider the total energy of a configuration $s$ as independent contributions from each lattice point, which gives
\begin{equation}
E^s=\sum\limits_{i} E_i \;.
\label{eq:energy}
\end{equation}
The state--update rules are deterministic and defined in Appendix \ref{app:Notation}. Given a rule, the assumption of independent energy contribution implies that the system does not follow a Hamiltonian evolution, like energy conservation, as expected from a cellular automaton. This assumption is further justified according to the purpose of obtaining a physically inspired classification scheme. 

Given a rule, the probability for the occurrence of a given configuration $P(E^s)$ could be specifically derived, together with an energy profile that reflects lower energy for the most common configurations.
%The probability enters in the thermodynamics definitions in Appendix \ref{app:statistical}.
However, deriving the specific probabilities for each cell based on the set of rules is exactly what one attempts to avoid by using a thermodynamical approach.
%Indeed, this is what this work attempts by studying large cellular automata. Hence the following approach is preferred.

For the purpose of this classification, we define the probability that a cell is in state $E_i$ as the probability that, if one picks a random cell from the system and checks what state it is in, it is found in state $i$. This probability corresponds to $P(E)=\frac{n(E)}{N}$, 
with $N$ being the total number of cells of the system, and $n(E)$ the number of cells at energy $E$.
By further assuming that each microstate at a given energy is equally likely, the expression for entropy (\ref{eq:1}), can be rewritten as {$S=-N\sum P(E)\log P(E)$} (cf. Appendix \ref{app:statistical}).
Following this, by observing that in a two level system $n(1)=E^s$ and $n(0)=N-E^s$, the entropy corresponding to a configuration is,
\begin{equation}
S=-\left( {E^s} \log \frac{E^s}{N} + (N-E^s)\log \left( 1-\frac{E^s}{N} \right) \right).
\label{eq:7}
\end{equation}
To be noted that the above expression is based on the assumption that all states at the same energy are equally likely. This assumption is known as the principle of equal a priori probability and it is considered to be a postulate of statistical mechanics. It is by no means obvious that this principle holds for cellular automata as well. In fact, it is not true in general: obviously the configuration with $E^s = 0$ is much more common than other for many rules. However, this assumption is useful to statistically study the characteristics of cellular automata, and its validity within this context will be checked in Sect. \ref{sec:ne}.

Once $S$ and $E^s$ are established, temperature can be calculated using their familiar relation leading to (cf. Appendix \ref{app:statistical}),
\begin{equation}
\label{BoltzmannTemperature}
T%=\frac{1}{\log (1-E^s/N)- \log (E^s/N)}
 =\frac{1}{\log (\frac{N-E^s}{E^s})}.
\end{equation}
Thus, the heat capacity $C$ is,
\begin{equation}
C=\frac{\partial E^s}{\partial T}=\left(\frac{\partial T}{\partial E^s}\right)^{-1}=-\frac{N}{E^s(E^s-N)(\log (1-E^s/N)-\log (E^s/N))^2} \;,
\end{equation}
and the Helmholtz free energy $A$ is
\begin{equation}
\label{Boltzmann_Helmholtz}
A=E^s-TS=N\frac{\log\frac{N-E^s}{N}}{\log\frac{N-E^s}{E^s}}\;.
\end{equation}

Pressure could be calculated from the Helmholtz free energy in relation to a change of volume. 
%In this work, the closest thing to volume/area is the number of cells so $P=-\frac{\partial A}{\partial N}$, this leads to a complex expression for pressure with an asymptotic behaviour at $E^s=N/2 $, which is also negative over its whole domain.
%$$P=-\frac{N \log (1 - E^s/N) + \log (-1 + N/E^s) (E^s + (N - E^s) \log (1 - E^s/N))}{(N - E^s) \log ^2(-1 + N/E^s)}. $$
Alternatively, there is the option of deriving pressure in analogy to its microscopic definition, counting of the interactions at the boundaries. Namely, as the energy density of the boundaries, which is the energy of the boundary cells divided by the number of boundary cells $4\sqrt{N}-4$, so
\begin{equation}
\label{pressureden}
P=\sum_{i \in boundaries} \frac{E_i}{4 \sqrt{N}-4}.
\end{equation}
%This way of defining pressure is similar to the microscopic definition of pressure in a thermodynamical system, which takes into account the force that the molecules exert on the boundaries of the system. 

% For the sake of simplicity, from the next section on, the total energy of a configuration $E^s$ will be denoted by just $E$.
We will use  (\ref{eq:7}-\ref{pressureden})  as benchmark for the behaviour of cellular automata and compare these quantities to the ones derived from a partition function or in analogy with the ideal gas. 

\subsection{Partition Function}
%By having a value for entropy and energy one can calculate the temperature with Eq. (\ref{eq:2}). Thus, the partition function can be evaluated. 
The partition function of the binary cellular automaton with energy (\ref{eq:energy}) is derived in Appendix \ref{app:statistical} with the assumption of equal a priori probability, which holds true particularly for small energy variation. This leads to,
\begin{equation}
Z=\frac{e^{-N/T}(e^{\frac{N+1}{T}}-1)}{e^{1/T}-1},
\end{equation}
where $N$ is the number of cells in the lattice and $T$ the temperature. 

From the partition function it is possible to analytically derive the relevant thermodynamical quantities. Considering $\log (Z)=-N/T+\log (e^{\frac{N+1}{T}}-1)-\log (e^{1/T}-1)$, for large $N$ and $N \gg T$ the first and the second term in the expression cancel out, leading to $\log (Z)=-\log (e^{1/T}-1)$. Thus one can rewrite the aforementioned thermodynamical quantities as ,
\begin{equation}
\langle E^s \rangle =\frac{e^{1/T}}{e^{1/T}-1}, 
\end{equation}
\begin{equation}
C_Z=\frac{e^{1/T}}{(e^{1/T}-1)^2T^2}, 
\end{equation}
\begin{equation}
\label{Z_Helmholtz}
A_Z=T \log (e^{1/T}-1), 
\end{equation}
and similarly $S$ and $P$, 
\begin{equation}
S_Z=\frac{e^{1/T}}{T(e^{1/T}-1)}-\log (e^{1/T}-1)
\end{equation} and 
\begin{equation}
P_{Z}=\frac{\log \frac{N-2E^s}{E^s}}{(N-E^s) \log^2 \frac{N-E^s}{E^s}}-\frac{1}{(N-2E^s)\log\frac{N-E^s}{E^s}},
\end{equation}
which has been obtained using (\ref{BoltzmannTemperature}) for temperature.

\subsection{Ideal gas}
\label{sec:id}

%The next step would be to use established results from the ideal gas model to evaluate the thermodynamical properties of cellular automata.
With the interpretation of cells as point-like particles, volume and particle number are essentially equivalent (or proportional), both corresponding to $N$, leading to significant simplifications as shown in the quantities below.
In the framework of the ideal gas, energy can be derived from temperature as $E=\frac{3}{2} NT$, in the case of a two-dimensional system, this should become $E^s=NT_I$ that defines the ideal gas temperature $T_I$.

The entropy of an ideal gas is described by the Sackur-Tetrode equation. In this case, it is necessary to derive the two-dimensional Sackur-Tetrode equation (cf. App. C),
 Therefore,
\begin{equation}
\label{entropyeq}
S_I=N \log (E^s).
\end{equation}

The ideal gas temperature $T_I$ can be also obtained as $T_I=\frac{\partial E^s}{\partial S_I}=(\frac{\partial N \log E^s}{\partial E^s})^{-1}=E^s/N$. So,
\begin{equation}
\label{T_Ideal}
T_I=\frac{E^s}{N},
\end{equation}
in correspondence with the previous definition. Despite the assumptions made in the derivation of $S_I$ it still reproduces the expression $E^s=NT_I$. Similarly, the heat capacity $C_I=\frac{\partial E^s}{\partial T_I}$ becomes $C_I=N$, following the two-dimensional ideal gas. 

Using the ideal gas entropy (\ref{entropyeq}) and temperature (\ref{T_Ideal}), the Helmholtz free energy becomes
\begin{equation}
A_I=E^s-E^s\log (E^s)=NT_I-NT_I\log (NT_I).
\end{equation}
Thus, 
\begin{equation}
P_I=-\frac{\partial A_I}{\partial N}=T_I\log (NT_I).
\end{equation}

Another possibility for the calculation of temperature is to consider another physically inspired definition. $T_A$ is defined based on the activity of the cellular automaton, by analogy to the concept of the average kinetic energy of the molecules,
\begin{equation}
\label{ActivityTemperature}
 T_A=\frac{1}{N} \sum_{i=1}^{N} | E_{i,t+1} - E_{i,t} |.
 \end{equation}
% $$ a_i = \begin{cases} 1, & \mbox{if } \mbox{the state of the cell $i$ is different from the previous generations} \\ 0, & \mbox{if } not \end{cases}$$
That is, $T_A$ is the sum of all value changes in the automaton divided by the number of cells \cite{simon}.
The interpretation of temperature exclusively as the average kinetic energy of the molecules is valid only for an ideal gas model. We are going to use this definition to classify the rules where the ideal gas model can be applied to cellular automata and the assumptions of the ideal gas hold. Note also that we have assumed independent energy contributions from the active lattice points. This implies that those rules which update the state of the system compatibly with an independent particle evolution, will be interpreted as "ideal" within this scheme. This will help identifying and classifying the rules. 

%\pagebreak

\subsection{Nonequilibrium Thermodynamics}
A final approach to the problem of the thermodynamics of cellular automata is provided by the formalism of nonequilibrium thermodynamics, as developed by Prigogine~\cite{prigogine}.
This approach is based on the concept of local equilibrium. That is, the parts of a thermodynamic system outside of equilibrium are individually in local thermodynamical equilibrium. This allows to define intensive quantities as time and space-dependent, while extensive quantities are substituted by their relative densities, so $T(x,t)$, $P(x,t)$, $s(x,t)=S(x,t)/N$ and $e(x,t)=E(x,t)/N$.
This approach to the thermodynamics of the system assumes that the quantities do not depend on the gradients of the system. That is, the system is not far from equilibrium.

This work attempts to get an overview of the applicability of the postulate of local equilibrium for some cellular automata rules by partitioning the system into boxes and recursively compute the thermodynamical variables. Therefore, in this way justifying the approaches above based on the principle of equal a priori probability. Specifically, if the energy density has similar values across the partition, then the principle of equal a priori probability is valid.

There is a trade-off when choosing the number of boxes to partition the system into. A high number of boxes gives a better picture of locality but it is a problem in the fact that the quantities are not well defined for small boxes since the thermodynamical limit is not reached. Vice-versa, when there are too few boxes, each box has well-defined quantities but the picture of locality is less detailed. A good compromise is found by using 25 boxes so that each box has 6400 cells when using a cellular automaton of 160000 cells. The output consists of a set of local averages of the variables over the evolution. 

%The hope is to get some insight into the potential of the non-equilibrium thermodynamics formalism for cellular automata and eventually lay the groundwork for more detailed studies in the future.

\subsection{Classification}

% Regarding the temperature used in calculating the partition function, one could choose between $T$ (\ref{BoltzmannTemperature}), $T_I$ (\ref{T_Ideal}), and $T_A$ (\ref{ActivityTemperature}) which have all been tested with satisfactory results. The quantitative part of this project makes use of the temperature from the benchmark (\ref{BoltzmannTemperature}) since it is the most fundamental definition among the three.

We investigate the rules based on the approaches previously defined. Specifically, we look if the system can be reliably described using partition function to identify it as a system in thermodynamical equilibrium. At the same time, we consider if the system's activity corresponds to the ideal gas temperature. The rules which reach zero activity in a 'short time', less than 100 iterations, are classified as "dead". By using this classification technique one could restrict the pool of rules among which to choose for a study. If, for instance, one would want to study non-equilibrium behaviour, instead of having to choose among all 18150 rules, it would only need to choose among rules that fall on that end of the classification scheme.

To evaluate whether two thermodynamical quantities are giving comparable results we consider the coefficient of variation. The coefficient of variation is a very useful metric that indicates the sample standard deviation  $\sigma$ in relation to the sample mean $\mu$ and it is thus scale independent.
We calculate the coefficient of variation with respect to different definitions of the Helmholtz free energy for benchmark (\ref{Boltzmann_Helmholtz}) and partition function (\ref{Z_Helmholtz}),
\begin{equation}
 CV_A = \frac{\sigma(A(t)/A_Z(t))}{\mu(A(t)/A_Z(t)},
 \end{equation}
where $\sigma$ and $\mu$ are calculated among the set of the values for the first 40 time steps $t$. If this value is smaller than a certain threshold, the two values of the Helmholtz free energy are proportional, hence are equally descriptive, which means that the partition function is a good approach to the system. Therefore, we classify it to be an equilibrium cellular automaton. The reason behind the choice of the Helmholtz free energy is that it is the most direct metric from the partition function, whereas expected energy and heat capacity require derivatives.

To evaluate if a rule corresponds to an ideal gas or not, we consider the coefficient of variation of the ratio between the ideal temperature (\ref{T_Ideal}) and the activity temperature (\ref{ActivityTemperature}),
\begin{equation}
     CV_T = \frac{\sigma(T_A(t)/T_I(t))}{\mu(T_A(t)/T_I(t))},
\end{equation}
where $\sigma$ and $\mu$ are calculated among the set of the values for the first 25 time steps. If this value is smaller than a certain threshold, the two temperatures are proportional, hence the ideal gas temperature can be viewed as an average activity. Therefore, we classify it to be a cellular automata akin to the ideal gas.

To define and verify the stability of the classification scheme, we run every single rule over four settings (cf. Appendix \ref{app:Notation}):
\begin{itemize}
\item $400 \times 400$ cells cellular automata with periodic boundary conditions;
\item $400 \times 400$ cells cellular automata with closed boundary conditions;
\item $100 \times 100$ cells cellular automata with periodic boundary conditions;
\item $100 \times 100$ cells cellular automata with closed boundary conditions.
\end{itemize}
If different runs lead to a different classification we say that a rule has switched classes. 
The final classification thresholds are defined as the non-trivial thresholds that lead to the lowest number of class switches. These values are obtained through a grid search with a granularity of 0.01 for both thresholds.

%I would not be surprised if the reader was a bit confused at this point, therefore 
% To conclude the section, a summary of the approaches and main equations is provided in table \ref{table:1}.
% \begin{table}[H]

%     \centering
% \resizebox{\textwidth}{!}{\begin{tabular}{c|c|c|c|c|c|c}
%          &&&$\frac{\delta E^s}{\delta S}$&$\frac{\delta E^s}{\delta T}$&$A=E^s-TS$&$\frac{\delta A}{\delta N}$\\
%          &$E^s$&$S$&$T$&$C$&$A$&$P$  \\ \hline
%          Benchmark & $\sum_i E_i$ & $-\sum_s P_s \log P_s$& $\frac{1}{\log (\frac{N-E^s}{E^s})}$ & $\frac{\delta E^s}{\delta T}$ &$N\frac{\log\frac{N-E^s}{N}}{\log\frac{N-E^s}{E^s}}$ & $\sum_{i \in boundaries} \frac{E_i}{4 \sqrt{N}-4}$ \\ 
%          Partition Function & $\sum_i E_i$  &$T \frac{\delta \log Z}{\delta T}+\log(z)$ & $\frac{1}{\log (\frac{N-E^s}{E^s})}$ &$\frac{1}{T^2} \frac{\delta^2 \log (Z)}{\delta^2 (\frac{1}{T})}$ &$-T \log Z$ & $P_{Z}=\frac{\log \frac{N-2E^s}{E^s}}{(N-E^s) \log^2 \frac{N-E^s}{E^s}}-\frac{1}{(N-2E^s)\log\frac{N-E^s}{E^s}}$\\ 
%          Ideal gas &  $\sum_i E_i$  & $N \log (E^s)$ &$E^s/N$ &$N$ &$E^s-E^s\log (E^s)$ &$T\log (NT)$ 
%     \end{tabular}}
%     \caption{Summary table with the definitions of the thermodynamical quantities as used in this study.}
%     \label{table:1}
% \end{table}

\section{Results}

We initialize each cellular automata with a random sample of total energy $E^s = 0.1 N$ and let it evolve for 100 iterations.
One important remark to make is that in some cases this might not reach the asymptotic state. However, there is no need to analyse the asymptotic state since the nature of the transient itself might be of interest and therefore merit a classification.Furthermore, a considerable amount of rules show zero activity before reaching the 100th iteration. That is, cellular automata often "freeze out". This is not surprising, since a large number of rules will have many deaths respect to few birth conditions or vice versa. For example, the rule with death condition for 1 to 7 neighbours and birth condition only with 8 alive neighbours will invariably die, unless it were fully populated and with periodic boundary conditions. Depending on the boundary conditions and the size, this number varies slightly. Generally, it oscillates around 3500, which is close to one-fifth of all rules taken into account. We classify these rules separately as "dead".

\begin{figure}[H]
    \centering
    \includegraphics[height=10cm]{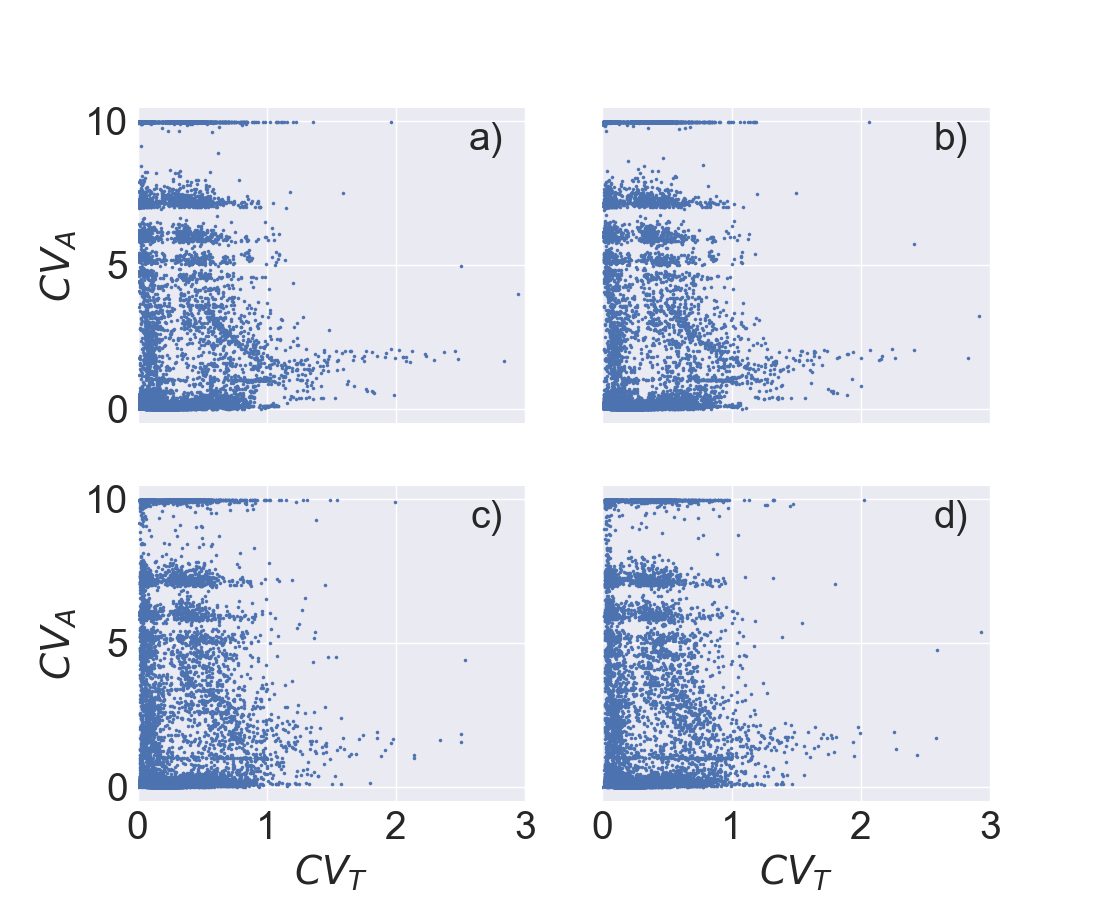}
    \caption{Scatter plots of $CV_A$ against $CV_T$. Each point is one of the totalistic binary cellular automata rules. (a) $400 \times 400$ cellular automata with periodic boundary conditions; (b) $400 \times 400$ cellular automata with closed boundary conditions; (c) $100 \times 100$ cellular automata with periodic boundary conditions; (d) $100 \times 100$ cellular automata with closed boundary conditions.}
    \label{fig:classes}
\end{figure}

Figure \ref{fig:classes} displays the distribution of $CV_A$ in scatter plots against $CV_T$ for the different sizes and boundary conditions. It is possible to appreciate the essentially similar results in the four cases, comparing between small and big sizes and between closed and open boundaries. Generally, we have found that cellular automata show the same or similar thermodynamical behaviour with periodic and closed boundary conditions and with variations in size. This shows that the thermodynamic quantities defined are often valid and fairly stable and that the random initialization doesn't affect the result in a relevant way.

\begin{figure}[H]
    \centering
    \includegraphics[height=8cm]{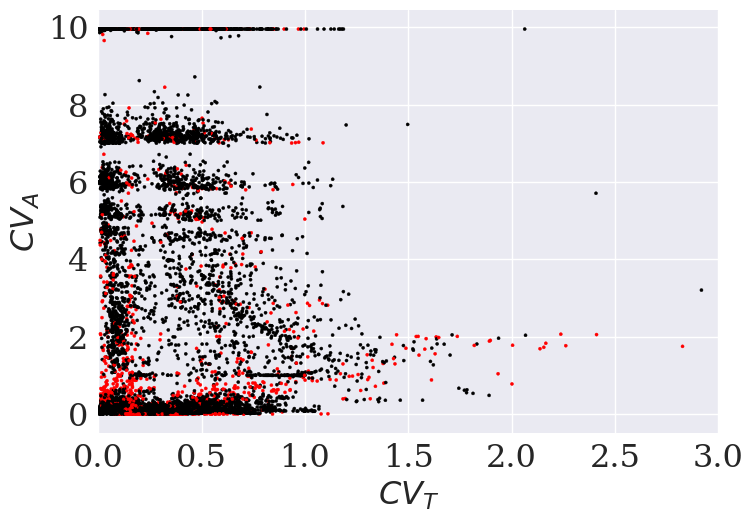}
    \caption{Scatter plots of $CV_A$ against $CV_T$ for cellular automata of size $400 \times 400$ with closed boundary conditions with a highlight on the rules that switched classes when open boundaries are applied, based on the thresholds of 0.16 for $CV_T$ and 0.73 for $CV_A$.}
    \label{fig: switch}
\end{figure}

\begin{figure}[H]
    \centering
    \includegraphics[height=8cm]{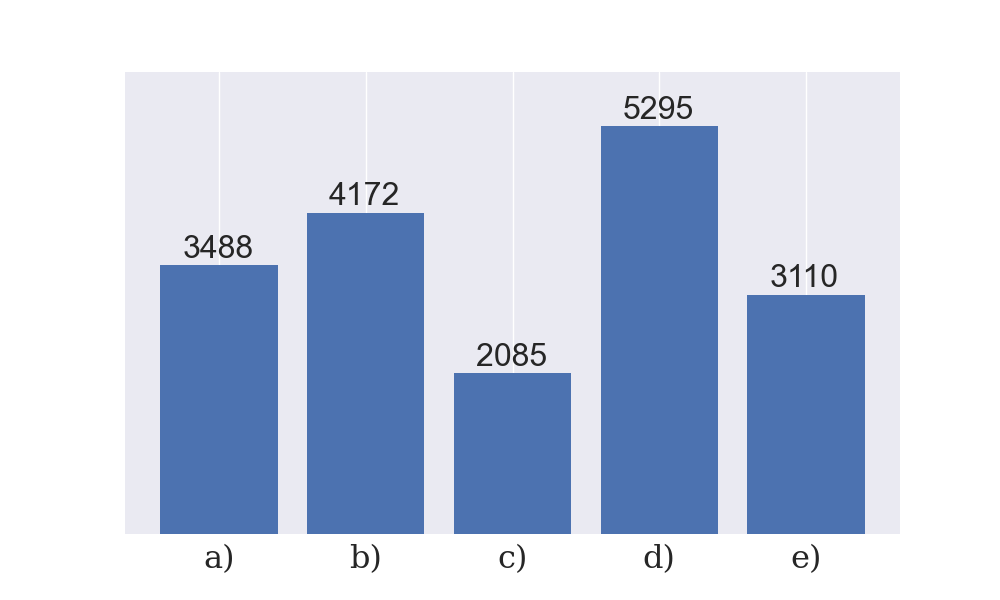}
    \caption{Result of the classification for $400 \times 400$ cellular automata with closed boundary conditions. The resulting classes are (a) dead, corresponding to inactivity before the $100$th iteration; (b) ideal gas in equilibrium; (c) ideal gas outside of equilibrium; (d) non-ideal gas in equilibrium; (e) non-ideal gas outside of equilibrium.}
    \label{fig:class_hist}
\end{figure}

Following these results, the thresholds that minimize the number of rule switches have been found to be 0.16 for $CV_T$, which resulted in 1772 class switches and 0.73 for $CV_A$,  which resulted in 1113 class switches. The total number of rule switches is 2819 which is less than the sum of the individual switches since some rules switch in both cases. We show a graph with a highlight on the rules that switched in figure \ref{fig: switch}. The relative small number of class switches over the several conditions testifies to the robustness of the chosen procedure and the correctness of the assumptions for the purpose of classification.
Figure \ref{fig:class_hist} shows an histogram of the final classification for each rule, based on $400 \times 400$ cellular automata with closed boundary conditions.

\begin{figure}[H]
    \centering
    \includegraphics[height=5.5cm]{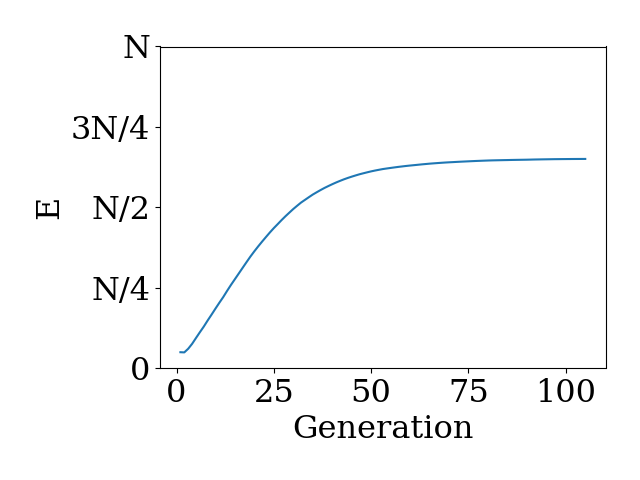}
    \includegraphics[height=5.5cm]{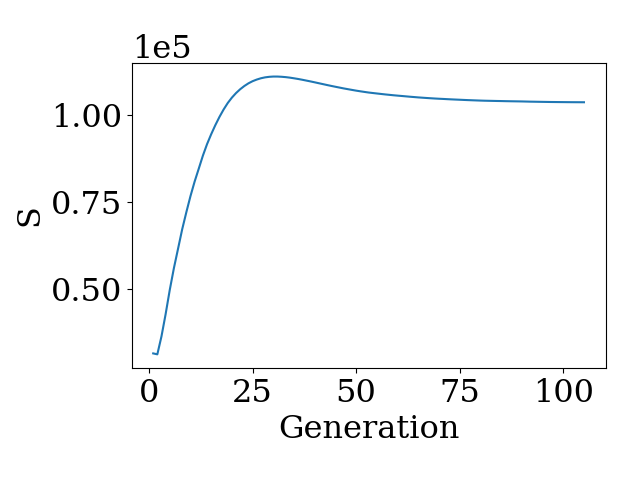}
    \includegraphics[height=5.5cm]{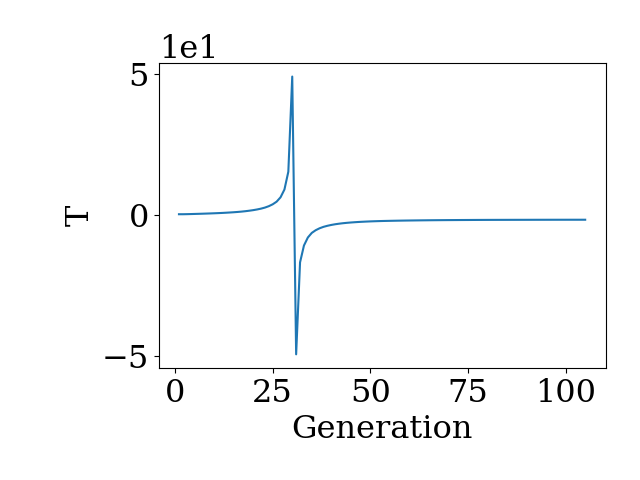}
    \includegraphics[height=5.5cm]{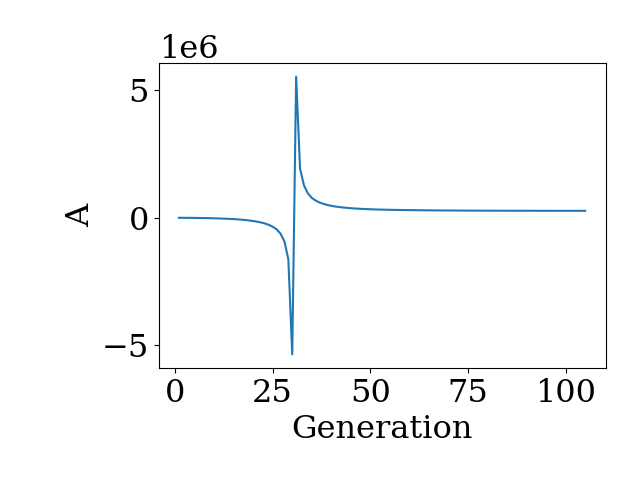}
    \includegraphics[height=5.5cm]{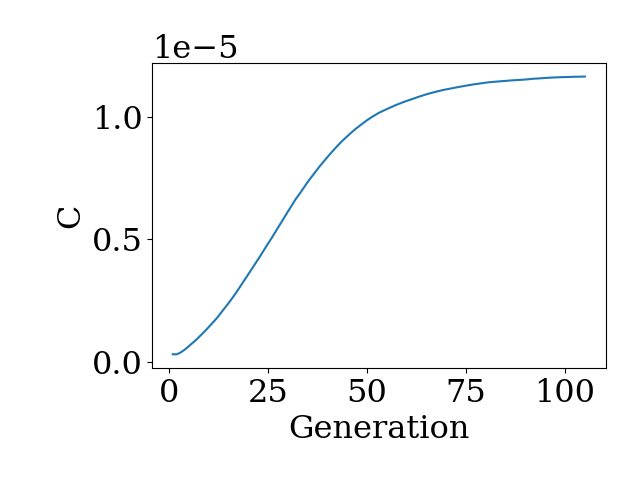}
    \includegraphics[height=5.5cm]{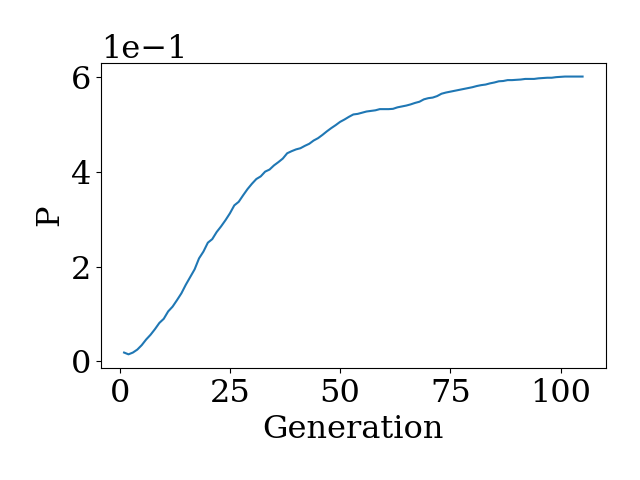}
    \caption{The figure shows the energy, entropy, temperature, Helmholtz free energy, heat capacity and pressure as defined in section \ref{sec:bm} for a cellular automaton with 160000 cells evolving according to to the rule with $B = \{3\}$ and $D = \{0,1,7,8\}$ (Non-ideal cellular automaton outside of equilibrium)}
    \label{fig: benchmark}
\end{figure}

\begin{figure}[H]
    \centering
    \includegraphics[height=5.5cm]{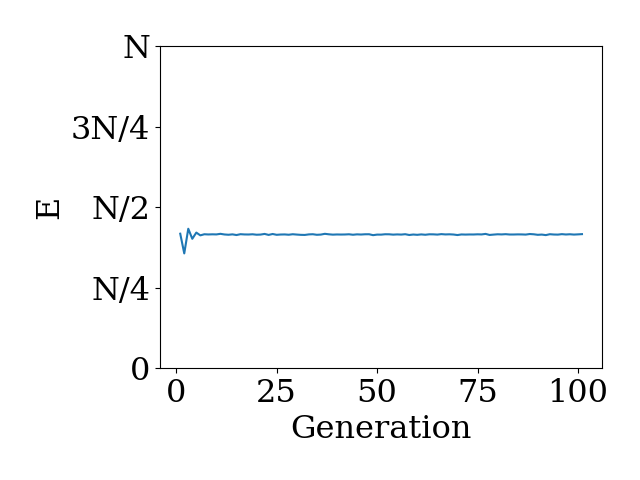}
    \includegraphics[height=5.5cm]{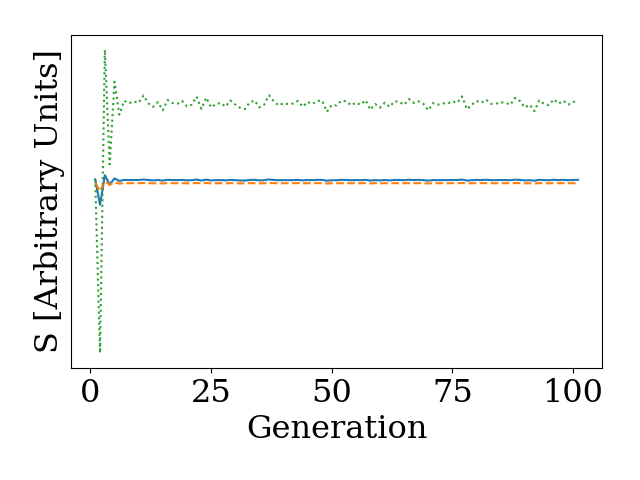}
    \includegraphics[height=5.5cm]{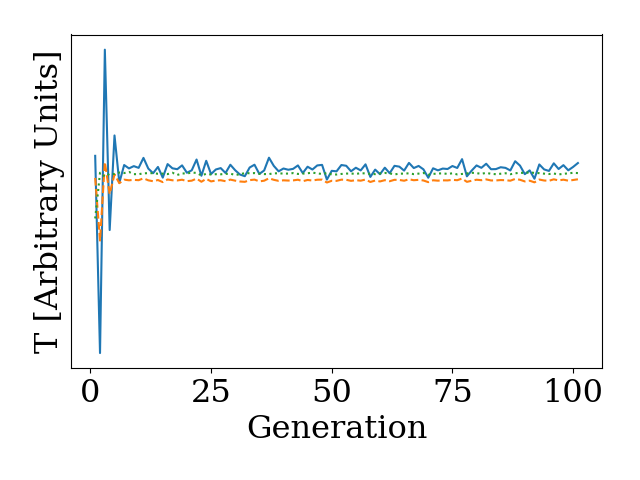}
    \includegraphics[height=5.5cm]{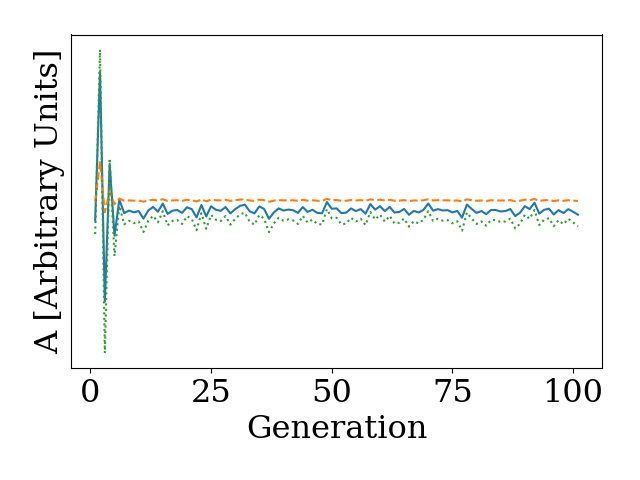}
    \includegraphics[height=5.5cm]{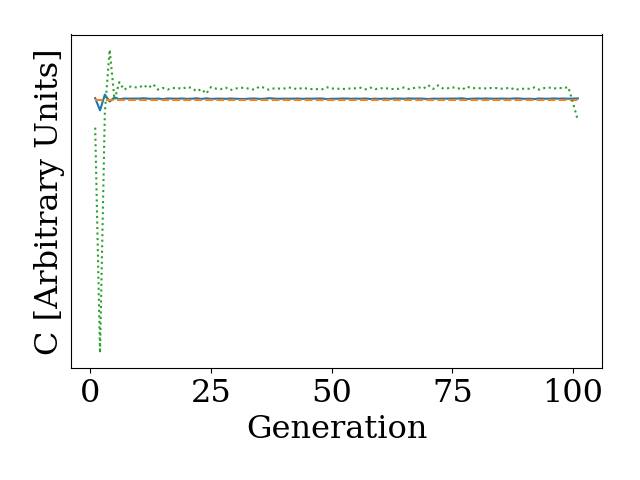}
    \includegraphics[height=5.5cm]{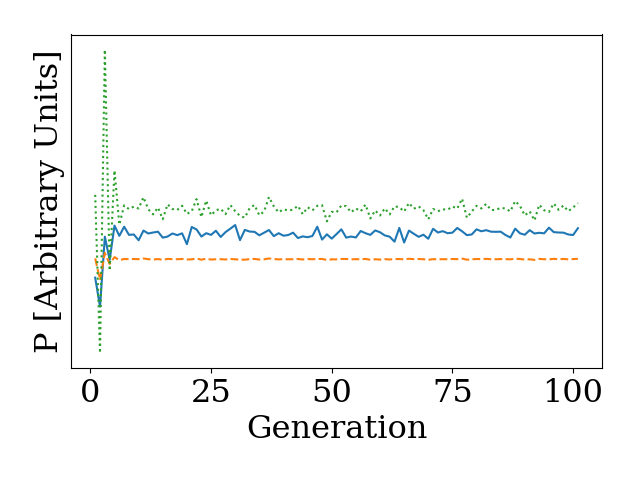}
      \caption{
      Results from indicated quantities calculated from the benchmark (solid line), the Ideal gas (dashed line) and the partition function (dotted line) evolving according to the rule with $B = \{1,3,6,7\}$ and $D = \{0,2,4,5\}$ (ideal cellular automaton in equilibrium). In the case of the temperature, the dotted line represents the average activity. All values have been normalized to the same starting point by multiplying the values with the ratio between the first value of the series considered and the first value of the benchmark series.}
    \label{fig: A}
\end{figure}

\begin{figure}[H]
    \centering
    \includegraphics[height=5.5cm]{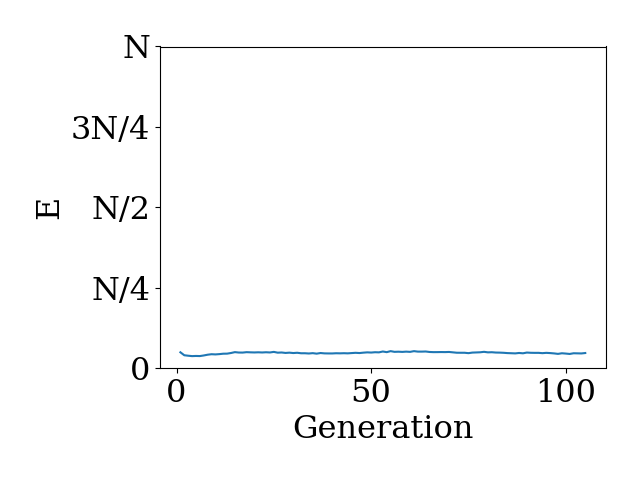}
    \includegraphics[height=5.5cm]{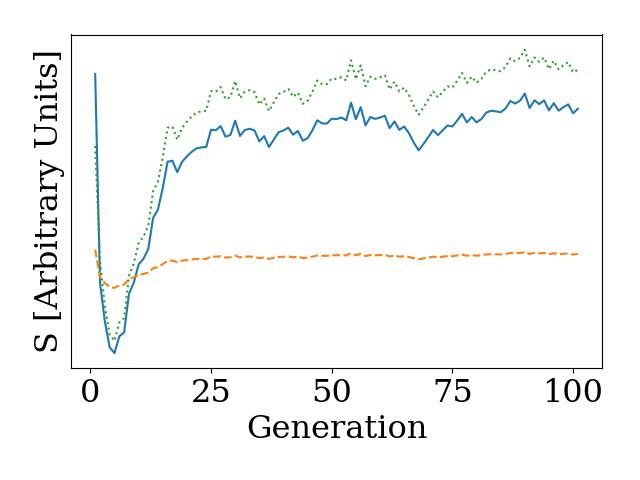}
    \includegraphics[height=5.5cm]{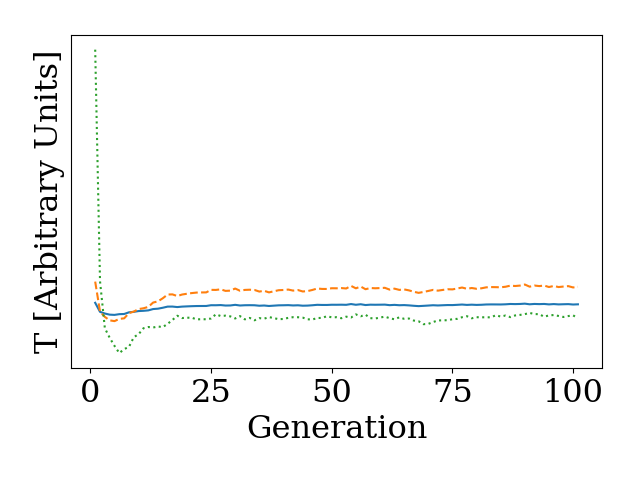}
    \includegraphics[height=5.5cm]{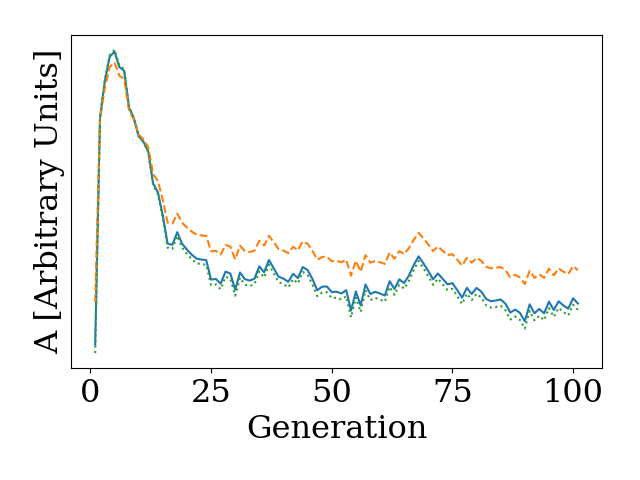}
    \includegraphics[height=5.5cm]{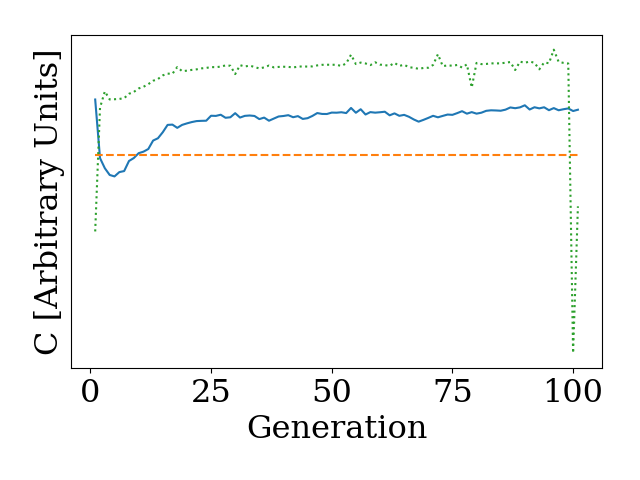}
    \includegraphics[height=5.5cm]{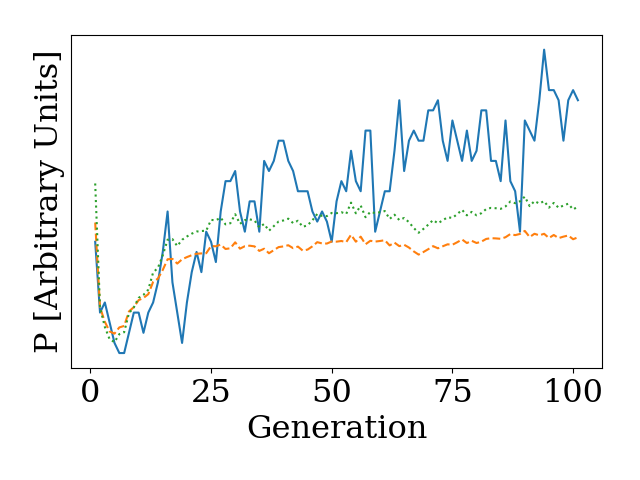}
      \caption{Results from indicated quantities calculated from the benchmark (solid line), the Ideal gas (dashed line) and the partition function (dotted line) evolving according to the rule with $B = \{3\}$ and $D = \{0,1,4,5,6,7,8\}$ (Game of Life). All values have been normalized to the same starting point by multiplying the values  with the ratio between the first value of the series considered and the first value of the benchmark series.}
    \label{fig: GOL}
\end{figure}

The classification results are scale-independent. However, the different quantities cannot be compared with each other directly, because of the different underlying assumptions in their derivation and of the different treatment of constants and units. In this work, constants have been disregarded due to the lack of a physical scale. A way of comparing the results is to normalize them to the same starting value as done in Figs. \ref{fig: benchmark} - \ref{fig: GOL}.
Generally, the results show some small scale fluctuations, which are a consequence of the finite and limited size of the automata. 
% \subsection{Benchmark}

The results obtained in Fig. \ref{fig: benchmark} are a direct consequence of the energy state $E^s$. This approach also correctly reproduces the transition to negative temperature states for an occupation $E^s>N/2$, where a system at a given temperature can evolve into a system at negative temperature. Of course this transition is not allowed in classical thermodynamics in the case of unbounded phase space and energy conservation \cite{book}. The transition is accessible to cellular automata because they do not necessarily obey energy conservation and the phase space provided is limited. The entropy $S$ of a two-state system (\ref{eq:7}) is maximized at $E^s=N/2$. Therefore, any move from that energy state will cause a decrease in entropy. If the energy increases further, occupying more sites, the temperature is negative.

Regarding the other quantities, the Helmholtz free energy is negative at positive temperature and vice versa for all rules and it is minimized when entropy is maximized, as expected. The heat capacity is always positive but has a somewhat unpredictable behaviour, sometimes stationary, sometimes increasing and sometimes decreasing with each time step.
The pressure is generally monotonously increasing with energy, which is a direct consequence of the fact that the energy tends to distribute evenly across the grid and it offers a weak but reassuring justification of the postulate of equal a priori probability. After all, the pressure (\ref{pressureden}) is defined just as a localized energy density. 
A more thorough justification of the principle of equal a priori probability is offered by the non-equilibrium approach described in section \ref{sec:ne}.

% \subsection{Ideal gas}

% In the case of the ideal gas approach, we start with the same definition of energy as in the previous section but define the other quantities as corrections to the known expressions for the ideal gas. This leads to an entropy (\ref{entropyeq}), which is monotonously increasing with respect to energy and temperature taking the form of average energy. 

% Regarding the other quantities, the Helmholtz free energy, given the simple relationships between the quantities, behaves in a well defined and expected way with respect to energy. This regular behaviour is observed also for the other approaches.
% The heat capacity is defined as constant. The ideal gas pressure is monotonously increasing with temperature and energy, which is a reasonable result.

% This exact behaviour is shown in \ref{fig: ideal} where the monotonously increasing energy leads to monotonously increasing entropy, temperature and pressure and a decreasing Helmholtz free energy towards equilibrium.

% \subsection{Partition Function}

The partition function approach is valid for equilibrium or near equilibrium cellular automata. It is fundamentally different from the other approaches since it is based on temperature instead of energy. It breaks down if negative temperatures are reached since $lim_{T \to 0^-} Z(T)=\infty$.

Overall, the metrics do not show very quick changes. Generally, these equilibrium rules either oscillate around equilibrium, as shown in Fig. \ref{fig: A} or reach an equilibrium state in a slow and controlled manner, i.e., with small gradients.
This is generally true for temperature, energy and entropy in the same way. The Helmholtz free energy is generally negative and evolves in opposite direction than the aforementioned quantities. Heat capacity tends to zero at low temperatures and to one at higher temperatures. An example of this approach is found in Fig. \ref{fig: A}, where all quantities oscillate around a stable value. The values for the ideal gas, partition function and benchmark are shown for all the quantities.

Finally, the reader might be interested in seeing the behaviour of the thermodynamical quantities of the well-known Game of Life. Figure \ref{fig: GOL} shows that.

% Regarding the pressure, $P_{Z}$ is negative for all energy and temperature values. Negative pressure can be hard to consider if one wants to interpret it as absolute pressure. 
% Negative pressure is a result that requires more theoretical interpretation for real physical systems \cite{1,2}. While it is possible to interpret negative pressure for some exotic rules as it is not ruled out by thermodynamics \cite{3}, it is not acceptable that it is the case for a whole class of rules and especially it seems unreasonable to be the case for equilibrium rules. However, when scaled to the same starting value as the benchmark pressure as shown in \ref{fig: A} and \ref{fig: GOL} it shows a reasonable behaviour. 

\subsection*{Local equilibrium}
\label{sec:ne}

Using the approach described in Sec. \ref{sec:ne}, we find that the energy is evenly distributed among the partitions for most rules.
This has been checked by calculating the spread between the energy density of the partition with the highest average value and the lowest average value when running the cellular automata for 100 time-steps. For $17684$ out of $18150$ rules this spread has been found to be below $0.1$ and in only $37$ rules above $0.3$. 
This provides a good validation of the principle of equal a priori probability. 

\begin{figure}[H]
    \centering
     \begin{subfigure}[t]{0.03\textwidth}
    \textbf{a)}
  \end{subfigure}
  \begin{subfigure}[t]{0.44\textwidth}
    \includegraphics[width=1\linewidth]{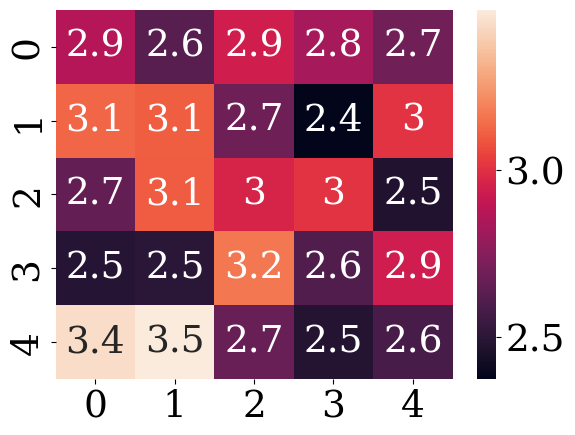}
  \end{subfigure}\hfill
  \begin{subfigure}[t]{0.03\textwidth}
    \textbf{b)}
  \end{subfigure}
  \begin{subfigure}[t]{0.455\textwidth}
    \includegraphics[width=1\linewidth]{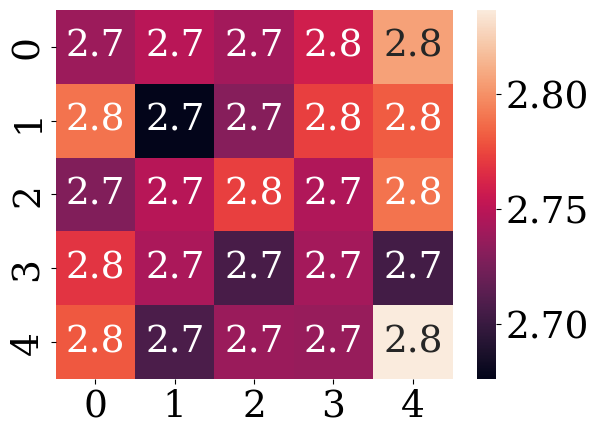}
  \end{subfigure}
    
    % \begin{subfigure}
    % \includegraphics[width=5cm]{D.png}
    
    % \end{subfigure}
    % \begin{subfigure}
    % \includegraphics[width=5cm]{A.png}
    % \end{subfigure}
    
    \caption{Temperature heatmap of a 160000 cells cellular automaton partitioned into 25 subdivisions and evolving according to the rule with $B = \{2\}$ and $D = \{0,4,6,8\}$ (a) and the rule with $B = \{1,3,6,7\}$ and $D = \{0,2,4,5\}$ (b).}
    \label{fig:noneq}
\end{figure}

However, for it is found that large number rules have a sizeable deviation of temperature among the partitions, an example is shown in Fig. \ref{fig:noneq}. This means that some rules deviate from thermal equilibrium, showing gradients and localities in the temperature distribution. We classify these cases as "non--equilibrium". 
Sometimes, when the energy of the automaton is close to $E^s=N/2$, the measures show some boxes at positive temperature and some boxes at negative temperature, which is expected, since the automaton is close to the temperature transition. 
Since the partition function breaks down for these rules, they have been classified as non-equilibrium even though they might a quite uniform negative temperature distribution. An example of the temperature distribution for both a rule showing equilibrium and a non-ideal far from equilibrium rule are shown in Fig \ref{fig:noneq}. The same figure is intended as a comparison between the spatial distribution of temperature in two extreme cases. 

% \begin{figure}[H]
%     \centering
%     \includegraphics[height=5.5cm]{[3, 5, 7]-[4, 0, 6, 1]_Energy.png}
%     \includegraphics[height=5.5cm]{[3, 5, 7]-[4, 0, 6, 1]_Entropy.png}
%     \includegraphics[height=5.5cm]{[3, 5, 7]-[4, 0, 6, 1]_Temp.png}
%     \includegraphics[height=5.5cm]{[3, 5, 7]-[4, 0, 6, 1]_Helmholtz.png}
%     \includegraphics[height=5.5cm]{[3, 5, 7]-[4, 0, 6, 1]_Heat capacity.png}
%     \includegraphics[height=5.5cm]{[3, 5, 7]-[4, 0, 6, 1]_Pressure.png}
%     \caption{The figure shows the ideal gas metrics of energy, entropy, temperature, Helmholtz free energy, heat capacity and pressure from section 2.2 of a cellular automaton with 160000 cells evolving according to the rule 357-0146 (Ideal cellular automaton outside of equilibrium)}
%     \label{fig: ideal}
% \end{figure}

\section{Conclusion}

The focus of this paper has been the exploration of the application of basic concepts of equilibrium statistical mechanics to two-dimensional cellular automata in order to provide a classification scheme.
%Most results come together comprehensively in both analytical derivations and numerical implementations. This helped to better comprehend cellular automata from a thermodynamics perspective.
The principle of equal a priori probability was verified to hold according to our definitions when tested using a local equilibrium approach for this purpose.
%Both show a fairly even energy density distribution, with the former being a stronger clue than the latter.
The quantities that have been defined successfully could be useful for the simulation and modelling of statistical mechanical systems and the understanding of the equilibrium properties of simple cellular automata.

The development of a physically derived quantitative classification scheme for two-dimensional cellular automata has shown to be robust for $84.45\%$ of the rules, with the rest being between classes.
Possibly, this work provides a step forward into more rigorous and quantitative classification systems for cellular automata with an unperturbed evolution from an initial density. Generally, the classification scheme used can be applied to the prediction of the behaviour of any model system that evolves according to local interactions, by individuating the cellular automata rule that corresponds to the interactions that govern it.

A natural next step for this work could be further investigate the application to the non-equilibrium formalism. Non-equilibrium thermodynamics methods could similarly be used to study every totalistic rule in this study instead of individual cases. Eventually the thermodynamical properties of systems of connected cellular automata, subject to different sets of rules and study the non-equilibrium processes associated with them (e.g. heat flow and diffusion). 
% Alternatively, a study could be done into making all the quantities coherent and using first principles to assign relative units to them, incorporating a more rigorous expansion towards the mathematical side.

%%%%%%%%%%%%%%%%%%%%%%%%%%%%%%%%%%%%%%%
\appendix
% \section{}
% The way the cellular automata is implemented is through the convolution of a neighbourhood matrix along the cellular automata state matrix. In Python this is done as follows for closed boundary conditions:
% \begin{lstlisting}[language=Python]
% from scipy import signal
% import numpy as np
% NEWmatrix = np.zeros((w,h),dtype= np.int8)
% kernel = np.ones((3, 3), dtype= np.int8) #Neighourhood
% kernel[1, 1] = 0 #Reduced moore neighbourhood
% k=signal.convolve(StateMatrix, kernel, mode='same')
% for i in range(w):
%         for j in range(h):
%             n = k[i,j] 
%             if n in B:                
%                 NEWmatrix[i][j] = 1                    #birth
%             elif n in D:         
%                 NEWmatrix[i][j] = 0                    #death
%             else:                               
%                 NEWmatrix[i][j] = StateMatrix[i][j]  #unaffected
% \end{lstlisting}
% This implementation is significantly faster than the naive for loop algorithm.

\section{Notation, neighbourhoods and boundaries}
\label{app:Notation}

This work analyses the behaviour of additive rules, so rules depend only on the total values of the cells in a neighbourhood.
The total number of possible additive rules, which considers the possibility of leaving the cell in its previous state is $3^{n+1}$ where $n$ is the number of cells in the neighbourhood. The base is three because each configuration of the $n$ cells of the neighbourhood can lead to three outputs, dead, alive or unchanged. The exponent is $n+1$  because the possibility of zero active neighbours is counted too. 
% \begin{comment}
% The notation to distinguish between the different rules is birth-death, which indicates the number of active sites needed for the activation and the deactivation of a cell. The numbers that are not included in the notation are the ones that leave the cell in its previous state.
% For example, if a cell is activated at 1 and 2 active neighbours and deactivated at 7 and 8, the rule will be described by 12-78. Recall that these numbers represent the total number of active neighbours and are independent of the position of the cells as long as they are in the neighbourhood. Specifically, using the notation of (\ref{below}) below, 1 and 2 would belong to $B$, while 7 and 8 would belong to $D$.
% \end{comment}
% The reason why only additive rules are studied is that this selection restricts the study enough for a comprehensive analysis that fits the temporal scope of this paper.
%Ideally, given enough time, one would be interested in analysing every kind of cellular automata rule.

Considering $w$ and $h$ the width and the height of the cellular automata (i.e. the number of cells per side) and $t$ the generation or time, the cellular automata at $t+1$ can be described as a function of the cellular automata at $t$ as the  $w \times h$ matrix, whose components are defined by the following expression
% $$C_{t+1}=E_{x,y,t+1}, \quad x \in \{1,2,...,w\}, \quad y \in \{1,2,...,h\} $$
$$E_{i}(t+1)=E_{\{x,y\}}(t+1), \quad x \in \{1,2,...,w\}, \quad y \in \{1,2,...,h\} $$
where $E_i$ is the state of the cell $i$, either $1$ or $0$.
If $B$ is the set of the number of sites that lead to the activation of the cell and $D$ is the set of number of sites needed for the death of a cell, satisfying the requirement $$B,D \subset \{0,1,...,8\}: B\cap D =\{ \emptyset \},$$
then $E_{i}(t+1)$ is defined as follows:
\begin{equation}\label{below}
		E_{i}(t+1) = \begin{cases} 1, & \mbox{if } \sum\limits_{i,j \in  Neighbourhood} E_{\{x+i,y+j\}}(t)-E_{\{x,y\}}(t) \; \in B \\
		0, & \mbox{if } \sum\limits_{i,j \in Neighbourhood} E_{\{x+i,y+j\}}(t)-E_{\{x,y\}}(t) \; \in D\\
		E_{\{x,y\}}(t)  & \mbox{otherwise} \end{cases}
\end{equation}
In this work, the focus is mainly in the reduced Moore neighbourhood, corresponding to all cells next to the central cell, including the diagonal and excluding the central cell.
% The reason behind this is that the Moore neighbourhood is a very common neighbourhood choice, and the limited scope of the work doesn't allow for the testing of several neighbourhoods.
With this kind of neighbourhood, the total number of rules is $3^9=19683$. We now exclude the rules that trivially do not lead to any new birth or any new death. These rules have a trivial outcome of full population or full starvation. Considering these corrections, the number of non-trivial rules is 18150 since the number of rules taken away is $3 \sum_{i=1}^9 {9 \choose i}=1533$.

Two kinds of boundary conditions are studied, closed and periodic. Periodic boundary conditions lead to the state of a cell at a boundary influencing the state of the cells at the opposite boundary (i.e. a torus).
Closed boundary conditions do not put value constraints on the cells at the boundaries, but limit their neighbourhood. That is, for a reduced Moore neighbourhood, a cell at the side will have five neighbours and a cell at a corner will only have three. 

\section{Derivation of partition function of binary cellular automata}
\label{app:statistical}
%Two fundamental concepts are that of microstate, which maps the configuration of the individual components of the system, and that of macrostate, which are the properties of the system as a whole, such as temperature and energy. Statistical mechanics has been successful in the study of the thermodynamics of mechanical systems such as gases, for which it is fundamentally impossible to solve the equations of motion for every particle involved.
In the case of cellular automata,
%it is possible to study the state of every single cell at any step. However, 
an approach based on statistical mechanics can simplify the study of the behaviour of large cellular automata in the same way it simplifies the study of gases and other physical systems.
This work covers various measures from statistical mechanics applied to cellular automata and compares them to benchmark quantities to derive a classification scheme.
%Therefore, a review of the use of these same concepts in thermodynamics is useful. 
Note that in this work the Boltzmann constant is kept implicit due to the lack of a natural physical scale.%  The reason behind this choice is that the interest of this paper lies in the relation between the thermodynamical variables in cellular automata, which are not defined by a natural physical scale such as the Boltzmann constant.

Entropy $S$  is at the core of many first principle definitions in thermodynamics, hence being of fundamental importance in evaluating thermodynamical properties of cellular automata. It is important to distinguish between the different uses of entropy as a measure of complexity since the many definitions coming from both physics and information theory can yield different results~ \cite{phase,complexity}.
One of the definitions of entropy used in statistical mechanics is the so-called Gibbs Entropy $S$, 
%which is the negative of the expected value of the logarithm of the probability $P_s$  that a system is in the microstate $s$\cite{book}, 
\begin{equation} \label{eq:1}
S=-\sum_s P_s \log P_s \;.
\end{equation}
Using this, the temperature can be defined as the partial derivative of the energy $E^s$ of the microstate $s$ with respect to entropy $S$, 
\begin{equation} 
\label{eq:2}
T= \left(\frac{\partial E^s}{\partial S}\right )_{\tilde N},
\end{equation}where the other state variables $\tilde N$ (such as the volume and particle number) are kept constant\cite{book}. 

For a system in thermal equilibrium with a heat bath (or reservoir), the total energy of the system and the reservoir is $U_0$, the energy of the system is $E^s$ and the energy of the reservoir is the difference $U_0 - E^s$.
The probability that the system is in state $s$ with energy $E^s$ is given, for a very large reservoir in thermodynamical equilibrium, by\cite{book}
\begin{equation}\label{eq:3}
P(E^s)=\frac{e^{-E^s/T}}{Z},
\end{equation}
with \begin{equation}\label{eq:4full}
Z=Z(T)=\sum_s e^{-E^s/T}=\sum_{i=0}^N d_i e^{-i/T},
\end{equation}
where $Z$ is the partition function of the system, and $d_i$ the number of configurations at energy $i$.
Since $d_i$ is a binomial coefficient of large numbers it is difficult to implement computationally (in the case of a grid $N = 400\times 400$ $d_i$ spans between $d_0 = d_N = 1$ and $d_{N/2} \approx 10^{48162}$). Therefore, keeping in mind we are interested at the partition function as benchmark for equilibrium, we consider the case where the energy variation is small, which is also the case for most of the rules after an initial transient period. In this case, the degeneracy can be approximated with a constant $d_i \approx d$, and therefore ignored, obtaining
\begin{equation}\label{eq:4}
Z(T) \approx \sum_{i=0}^N e^{-i/T}.
\end{equation}
%$Z$ is very useful in defining the thermodynamical properties of a system.
Through the partition function, one can derive the average energy, the Helmholtz free energy and in turn the pressure, the energy fluctuations and the heat capacity of the system at equilibrium.

%The partition function can be useful in evaluating a whole set of other quantities, at equilibrium.
In this case, the Helmholtz free energy is  $A_Z=-T \log (Z)$,  which can be compared to the common definition $A= E^s-TS$.  The average energy is $\langle E^s \rangle =-\frac{\partial \log (Z)}{\partial \frac{1}{T}}= T^2 \frac{\partial \log Z}{\partial T}.$ 
The variance of the energy is calculated as $(\Delta E^s)^2=\frac{\partial^2 \log (Z)}{\partial^2 (\frac{1}{T})}$ and it is useful for the heat capacity  $C_Z=\frac{(\partial E^s)^2}{T^2}$, which can be compared to the common definition $C=\frac{\partial E^s}{\partial T} $\cite{book}.
Entropy can also be calculated from the partition function starting from (\ref{eq:1}) and (\ref{eq:3}), this leads to $S_Z=-\sum_s P_s  \cdot ( -E^s/T - \log (Z))= \frac{\langle E^s \rangle}{T}+\log(Z)$.

It can also be shown (i.e. by induction) that $Z$ of Eq. (\ref{eq:4}) corresponds to 
\begin{equation}
Z=\frac{e^{-N/T}(e^{\frac{N+1}{T}}-1)}{e^{1/T}-1},
\end{equation}
\begin{proof}
\begin{align*}
N=1 \Longrightarrow Z=1+e^{-1/T}  &, \quad  \quad Z=\frac{e^{-1/T}(e^{2/T}-1)}{e^{1/T}-1}=\frac{(e^{1/T}-1)(e^{-1/T}+1)}{e^{1/T}-1} =1+e^{-1/T}\\
N=K \Longrightarrow Z=\sum_{i=0}^K e^{-i/T} &, \quad  \quad  Z=\frac{e^{-K/T}(e^{\frac{K+1}{T}}-1)}{e^{1/T}-1}\\
N=K+1 \Longrightarrow Z=\sum_{i=0}^{K+1} e^{-i/T} &, \quad  \quad  Z=\frac{e^{-K/T}(e^{\frac{K+1}{T}}-1)}{e^{1/T}-1} + e^{-(K+1)/T}\\
    																								&	\quad  \quad	\quad							=\frac{e^{-K/T}(e^{\frac{K+1}{T}}-1)+e^{-(K+1)/T}(e^{1/T}-1)}{e^{1/T}-1}\\
                                                                                                       &   \quad  \quad  \quad                            =\frac{e^{1/T}-e^{-(K+1)/T}}{e^{1/T}-1} \\
                                                                                                          &    \quad  \quad  \quad                        =\frac{e^{-(K+1)/T}(e^{(K+2)/T}-1)}{e^{1/T}-1} \qedhere
\end{align*}
\end{proof}

\section{Sackur-Tetrode equation for two--dimensional system}
\label{app:Sackur}
The entropy of an ideal gas is described by the Sackur-Tetrode equation as $$S_I=k_BN \log \left(\frac{V}{N} \left(\frac{4\pi m}{3 h ^2} \frac{E^s}{N}\right)^{3/2}\right)+\frac{5}{2} N,$$ with $m$ being the mass, $h$ the Planck constant and $k_B$ the Boltzmann constant.
This is obtained starting from Boltzmann's entropy formula, $S_I=k_B \log (\Omega)$, where $\Omega$ is the multiplicity or number of microstates accessible to the system.
From equation 2.40 in \cite{Schroeder}, a three dimensional gas has multiplicity $\Omega \approx \frac{V^N(2\pi m E^s)^{3N/2}}{h^{3N}N!(3N/2)!}$, where the factor $3N$ accounts for dimensions. Accordingly, to find the two-dimensional multiplicity, one can substitute $V$ with $A$ (the area) and $3N$ with $2N$, leading to $\Omega \approx \frac{(2\pi A  m E^s)^N}{ (N!)^2 h^{2N}} $. Using Stirling's approximation, $N! \approx \sqrt{2 \pi N}N^N e^{-N}$ one gets $\Omega \approx \frac{(2\pi A  m E^s)^N}{2 \pi N^{2N+1} e^{-2N} h^{2N} }$. At large $N$, $2N+1 \approx 2N$ and $(2\pi)^{N-1} \approx  (2\pi)^N$ . By using these approximations and taking the logarithm one gets the

In our treatment, the area and the particle number correspond both to the number of cells $N$, and the mass of a cell is not defined. The constants can be grouped as $b=\frac{2 \pi m}{h^2}$, which leads to $S_I=N \log \frac{E^s}{N}+N(2+ \log (b) )$, we can define $B\equiv 2+ \log (b)$,  leading to  $S_I=N( \log \frac{E^s}{N}+B)$.
Generally, one can notice that $B$ should be a large number. The ratio between the mass of the particles and the Planck constant squared becomes a very large number for all gases. In our treatment, it is enough to have $B$ large enough that the entropy is always positive. 
$B$ can be taken to be $B \approx \log (N)$ to make the entropy extensive.

This is an analytical conclusion and a confirmed computational result, the analytical result is shown as 
\begin{equation}
C_I=\frac{\partial E^s}{\partial T}=\frac{\partial E^s}{\frac{\partial E^s}{\partial S}}=\frac{\partial E^s}{\frac{\partial E^s}{\partial N      \log E^s}}=\frac{dE^s}{dE^s/N}=N. 
\end{equation}

We not that it is possible to derive an ideal gas law for cellular automata, namely finding an analytical expression for pressure, in different ways. 
A first naive approach could be to start from $P_IV=NT_I$, and by noticing that $V \propto N$, getting to an expression in the form $P_I \propto T_I$.  Indeed $P_I$ should be a monotonously increasing function of $T_I$. This work starts by using $P_I=-\frac{\partial A_I}{\partial V}$, which is interpreted as  $P_I=-\frac{\partial A_I}{\partial N}$. From here, one can use the Helmholtz free energy,  as it is commonly defined $A_I=E^s-T_IS_I$.

%%%%%%%%%%%%%%%%%%%%%%%%%%%%%%%%%%%%%%%%%%%%%%%%%%%%%%%%%%%%%%
\bibliographystyle{apsrev4-2}
\bibliography{bibliography.bib}

\end{document}